\documentclass[journal,12pt,onecolumn,draftclsnofoot]{IEEEtran}

\makeatletter
\def\ps@headings{%
\def\@oddhead{\mbox{}\scriptsize\rightmark \hfil \thepage}%
\def\@ adversarynhead{\scriptsize\thepage \hfil \leftmark\mbox{}}%
\def\@oddfoot{}%
\def\@ adversarynfoot{}}
\makeatother
\ifCLASSINFOpdf
\else

\fi

\usepackage{epsfig}
\usepackage{array}
\newcolumntype{L}[1]{>{\raggedright\let\newline\\\arraybackslash\hspace{0pt}}m{#1}}
\newcolumntype{C}[1]{>{\centering\let\newline\\\arraybackslash\hspace{0pt}}m{#1}}
\newcolumntype{R}[1]{>{\raggedleft\let\newline\\\arraybackslash\hspace{0pt}}m{#1}}
\usepackage{amsmath}
\usepackage{amsthm} 
\usepackage{mathrsfs}
\newcommand{\bc}{\begin{center}}
\newcommand{\ec}{\end{center}}
\newcommand{\be}{\begin{equation}}
\newcommand{\ee}{\end{equation}}
\usepackage{flexisym}
\usepackage{algorithmicx}
\usepackage{algorithm}
\usepackage{xcolor}
\usepackage{soul}
\usepackage{subcaption}
\usepackage{algpseudocode}

\newcounter{mytempeqncnt}
\usepackage{amsmath,amssymb}

\newcommand{\bnu}{\begin{enumerate}}
\newcommand{\enu}{\end{enumerate}}
\usepackage{cite}
\usepackage{url}
\usepackage{breqn}
\usepackage{mathtools, cuted}
\usepackage{lipsum}
\usepackage{mathtools}
\usepackage{amsmath}
\usepackage{blindtext, graphicx}
\usepackage{multicol,lipsum}
\usepackage{cuted}
\usepackage{color}

\usepackage[margin=0.75in]{geometry}
\usepackage{amssymb}
\usepackage{amsfonts}
\usepackage{amsmath}
\usepackage{stfloats}
\usepackage{blindtext}
\usepackage{lipsum}
\usepackage{cuted}
\newtheoremstyle{case}{}{}{}{}{}{:}{ }{}

\begin{document}
\title{Energy-Efficient UAV Relaying Robust Resource Allocation in Uncertain Adversarial Networks}
\author{Shakil Ahmed,~\IEEEmembership{Student Member,~IEEE}, Mostafa Zaman Chowdhury~\IEEEmembership{Senior Member,~IEEE}, Saifur Rahman Sabuj,~\IEEEmembership{Member,~IEEE}, Md Imtiajul Alam, and Yeong Min Jang,~\IEEEmembership{Member,~IEEE}
\thanks{S. Ahmed  is with the Department of Electrical and Computer Engineering, Iowa State University, USA (email: shakilahmed@ieee.org).} \thanks{M. Z. Chowdhury is with the Department of Electrical and Electronic Engineering, Khulna University of Engineering \& Technology (KUET), Bangladesh, (email: mzceee@ieee.org).} \thanks{S. R. Sabuj is with the Department of Electrical and Electronic Engineering, BRAC University, Bangladesh, (email: s.r.sabuj@ieee.org).} \thanks{M. I. Alam is with the Department of Education, Iowa State University, USA (email: imtiaj@iastate.edu )}, \thanks{Y. M. Jang is with the Department of Electronics Engineering, Kookmin University, Seoul, Korea, (email: yjang@kookmin.ac.kr).}}
\markboth{}
{}

\maketitle

\begin{abstract}
The mobile relaying technique is a critical enhancing technology in wireless communications due to a higher chance of supporting the remote user from the base station (BS) with better service quality.
This paper investigates energy-efficient (EE) mobile relaying networks mounted on the unmanned aerial
vehicle (UAV) while the unknown adversaries try to intercept the legitimate link.
We aim to optimize robust transmit power, both UAV and BS along, with relay hovering path, speed, and acceleration.
The BS sends legitimate information, which is forwarded to the user by the relay.
This procedure is defined as information-causality-constraint (ICC).
We jointly optimize the worst-case secrecy rate (WCSR) and UAV propulsion energy consumption (PEC)
for a finite time horizon.
We construct the BS-UAV, the UAV-user, and the UAV-adversary channel models.
We apply the UAV PEC considering UAV speed and acceleration.
We derive EE UAV relay-user maximization problem in the adversarial wireless networks at last.
While the problem is non-convex, we propose an iterative and sub-optimal algorithm
to optimize EE UAV relay with constraints, such as ICC, trajectory, speed, acceleration,
and transmit power.
First, we optimize both BS and UAV transmit power and hovering speed for known UAV path planning
and acceleration.
Using the optimal transmit power and speed, we obtain the optimal trajectory and acceleration.
We compare our algorithm with existing algorithms and demonstrate the improved EE UAV relaying
communication for our model.
\end{abstract}


\IEEEpeerreviewmaketitle

\section{Introduction}

\IEEEPARstart{R}{ecently}, unmanned aerial vehicles (UAVs) communication is an emerging example
to assist next-generation remote users with reliable connectivity \cite{Antim}.
The UAV communication system is less expensive than the terrestrial base station (BS) platform
due to its swift, dynamic, on-demand, flexible, and re-configurable features.
Moreover, the UAV relay is controllable. Due to its higher altitude, it often experiences
a significant line of sight (los) communication links.

The UAVs can be loosely classified \cite{YZeng_10} based on operation, such as aerial BS, relay, and
collecting information. When the BS is malfunctioning, UAV is deployed to serve as aerial BS
\cite{YZeng_10}.
Moreover, the UAVs also stay quasi-stationary on the serving area to support the nodes.
The scenarios, such as the (BS) offloading in hot spots and BS hardware limitations,
require a fast service recovery, and the UAV is an excellent choice \cite{5}.
If the BS is not available due to expensive installation costs in physically unreachable areas,
the UAVs are deployed as relays to increase the BS capacity.
Thus, the UAV relays are responsible for providing wireless connectivity for remote users
in adversarial environments, such as natural disaster recovery, military operation, rescue operation, etc.
Moreover, UAV relays can be deployed to collect or disseminate information \cite{6}, \cite{7}.
Collecting or disseminating information is vital in various domains, such as periodic sensing or
smart cities application.
This is because sensors' operational power is reduced if the UAVs fly over them to communicate,
which results in a more extended network lifetime.

The UAV relay has two categories, mobile and static. 
Like mobile UAV, the static UAV has a better chance of los. It is not dynamic and consumes more propulsion energy. Moreover, the static  UAV serves the only region, such as the football stadium, where UAV does not need to move \cite{new_1}.
However, mobile relaying has more advantages than static relaying, such as cost-effectiveness,
swift deployment, and serving an on-demand basis \cite{8}.
Due to the mobility of relay, it provides opportunities for improving the wireless network performance
by adjusting the relay location.
In recent years, research has been conducted on UAV relaying because of its range extension
capability \cite{31}~–~\cite{34}.
Moreover, the application of the UAV relaying increases the overall system performance.
Unfortunately, due to the limited mobility of nodes and back-haul techniques,
most of the conducted research on UAV relaying is static.

The authors in \cite{new_2} studied the throughput
analysis for mobile UAV relay. They achieved the optimal BS and UAV transmit power while designing the trajectory. However, their proposed model is limited due to 1) considering one known adversary and 2) not considering the UAV energy consumption. Though their algorithm optimized the UAV power to design trajectory, it did not design the optimal energy-saving trajectory. This is because UAV transmit power (up to few Watts) has a negligible effect than propulsion energy consumption (PEC). UAV PEC is typically up to a few hundred  KWatts.

We design mobile relaying communication to make the problem practical. Unfortunately, there are new challenges for UAV relaying communications.
Specifically, onboard power consumption during the finite time limits the UAV relay performance due to its fixed size.
The energy-efficient (EE) UAV communication, defining total communicated information bits normalized by UAV PEC \cite{Sha_1, Sha_2}, is an essential paramount feature.
Moreover, the UAV must forward information from BS to users by ensuring the physical layer security.
Additionally, the UAV has broadcast nature communication links, leading to substantial physical layer security concerns for uncertain adversaries.

Researchers are working on designing EE UAV networks broadly. However, it needs more attention
to secure the network.
For example, the authors in \cite{10}~–~\cite{11} designed EE UAV communication.
However, they did not consider the UAV onboard energy consumption and physical layer security.
Moreover, they considered straightforward UAV path planning.
A robust resource allocation to maximize the secrecy is studied in \cite{12} in adversaries' presence.
However, the authors did not consider UAV energy minimization.
In our previous work in \cite{17}, we optimize the UAV worst-case secrecy rate (WCSR) in adversarial
networks via resource allocation.
We proposed an algorithm that considers information-causality-constraint (ICC) while maximizing WCSR.

A limited theoretical analysis of UAV relaying security was studied in \cite{new_3, QWang_1}. 
In \cite{new_3}, the authors investigated the joint BS/UAV power and trajectory optimization.  
The system considers adversaries, which are partially known by the UAV.  In their investigation, the UAV is considered as the aerial base station. However, the authors proposed model is limited to serving close users. Moreover, they did not consider UAV PEC, which is a crucial paradigm to design trajectory.  The authors in \cite{QWang_1} investigated the UAV security, and their proposed networks had UAV relay, BS, user, and one adversary. Their proposed model is limited to the perfect location of adversaries to the UAV. This assumption is the limited application in real-world scenarios. They considered neither robust resource allocation nor UAV EE. The authors in \cite{new_3, QWang_1} analyzed the security aspect to maximize the throughput. However, their algorithms did not have an optimal trajectory and UAV EE for long-distance users since they did not investigate 1) UAV relaying security for unknown adversaries 2) designing an optimal trajectory for EE UAV. 

To simplify the proposed model, the authors in \cite{R_22} considered friendly jammer UAV, which did patrol with a limited trajectory in static nature. However, this kind of assumption makes the problem infeasible in certain conditions, such as if adversarial attacks are happening out of range of jammer UAV. Thus, the introduction of relaying (i.e., buffer) complicates the resource allocation design. They also introduced a multi-antenna jammer UAV to guarantee secure communication. However, this assumption may also not secure the communication if the adversary has a similar antenna design pattern. They also added jammer created artificial noise generated. This noise may create reasonable interference to legitimate users as well. In this paper, we relaxed some of the assumptions made by \cite{R_22}. For example, we consider UAV trajectory is dynamic, which ensures comfortable and secure resource allocation. However, we adopt a similar model in \cite{R_22} for channel state information, where the nodes can receive information based on the location. 

The authors in  \cite{R_25}  investigated the resource allocation algorithm for UAV communication systems with solar energy enabling sustainable communication services to multiple ground users. Eventually, they jointly designed the 3D aerial trajectory and the wireless resource allocation to maximize the system sum throughput over a given period. However, they do not address the information transmission algorithm in a secure environment. Though they have introduced Solar-Powered UAV, this can solve only the energy issue when there is required sunlight for the solar panel to be charged. However, our proposed model is a good fit for all areas where UAV requires limited power. We also have optimized the PEC by propping an iterative algorithm. 

The authors in \cite{R_23} examined the surveillance paradigm, which tracks dubious high-attitude communications. The authors introduced jammers to monitor suspicious communication.  By analyzing the UAV's adversarial communication in jammers' presence, they proposed a model to optimize the data package in lower power consumption constraints by the proposed selection policy. On the other hand, the authors in \cite{R_24} analyzed a model legitimate UAV monitors dubious UAVs. They investigated how to achieve the dubious flight information UAVs. As \cite{R_23}, they have also introduced a friendly jammer who can jam the dubious receiver to reduce the suspicious UAV data rate. Eventually, they proposed a model that can maximize eavesdropping success. However, none of \cite{R_23}-\cite{R_24} can address an environment where UAV works as a relay and eavesdropping on the ground adversaries. In this paper, we address the limitation mentioned above.

The system considers adversaries, which are partially known by the UAV. 
In their investigation, the UAV is considered as the aerial base station. However, the authors proposed model is limited to serving close-distance users. Moreover, they did not consider UAV PEC, which is a crucial paradigm to design trajectory. 
The authors investigated the UAV security, and their proposed networks had UAV relay, BS, user, and one adversary. Their proposed model is limited to the perfect location of adversaries to the UAV. This assumption has limited application in real-world scenarios.
They considered neither robust resource allocation nor UAV EC.
 However, their algorithms did not have an optimal trajectory and UAV EE for the long-distance user since they did not investigate 1) UAV relaying security for unknown adversaries, 2) designing an optimal trajectory for  EE UAV.

In \cite{14}, UAV relaying communication is studied, which helps to forward independent data
to different users.
The authors maximized the data volume and relay trajectory by using a simple algorithm.
In \cite{15}, the authors optimized the UAV flying path at a fixed altitude.
The authors in \cite{16} investigated the optimal UAV trajectory, considering the UAV onboard energy
for the energy-aware coverage path.
Their study considered a quad-rotor UAV measurement-based energy model, which was applied to aerial imaging.
Mobile UAV communication is studied in \cite{18} by assuming that the relay moves randomly,
which follows a specific mobility model.
They maximized the UAV mobile relay statistical characteristics via throughput.
The authors investigated throughput for UAV relaying networks in \cite{19}.
They achieved the minimum UAV transmit power and trajectory.

All of the above works consider UAV trajectory optimization.
There is still scope for research to design EE UAV communication and ICC while the adversaries try to hide in the wireless networks.
Adversaries often use artificial noise, which increases the wireless network noise level. This feature helps them hide their presence even when the user is close to the relay. Moreover, the adversaries collaborate, making their presence protect from legitimate nodes.
Also, most aircraft track optimization investigations are not studied for wireless networking purposes.

The above proposals and models aim to achieve optimal solutions on simplified algorithms.
Thus, we focus on developing a more real-world model and achieving a sub-optimal EE UAV relaying networks
using an iterative algorithm.
We design EE UAV mobile relaying via optimizing the UAV and the BS robust resource allocation,
which considers the joint WCSR and UAV PEC.
The best channel modeling is naturally achieved for the maximum throughput
if the UAV mobile relay stays fixed to the user's possible nearest location.
However, this scenario results in inefficient UAV PEC modeling due to the UAV hovering at zero speed
\cite{22}.
Thus, there must be an optimal trade-off between maximizing the average WCSR and also optimizing PEC.
Our main contributions in this paper are described as follows:

\begin{itemize}

\item We consider a scenario where the user receives data from the BS.
Due to the longer distance, there is no direct link between them.
Thus, UAV relaying is a promising solution to forward data to the user.
The system has uncertain adversaries who try to intercept the UAV-user link.
We achieve optimal to gain the average WCSR via optimizing joint UAV/BS transmit power and the UAV trajectory.
To achieve the EE, we employ the fixed-wing UAV PEC,
the function of speed and acceleration.
Based on this model, we formulate the EE UAV relay.
We did not write the channel model with blockage and non-line of sight (nlos) as we focus on our main contributions, such as applying information causality constraints and energy efficiency model effect in an adversarial environment. Moreover, we consider the rural area for our investigation. Being a rural area with fewer obstacles, we can safely assume the altitude of UAV is enough to avoid obstacles with possible los communication links. We also found enormous well-reputed papers considering the same type of channel model. However, very few of them are considering an uncertain adversarial environment. We have added one potential scenario for our proposed system model. 

\item We investigate EE UAV relaying maximization problem subject to ICC, UAV trajectory, speed,
acceleration, UAV/BS power constraints.
ICC is applied to capture the UAV broadcast communication from the BS.

\item Though we optimize UAV trajectory to achieve the optimal EE, we can also investigate WCSR with the optimal UAV trajectory. We have added the algorithm in Algorithm 2 in Section IV. Note that we have a power consumption model in the denominator of the original optimization problem. Since all the UAV trajectory variables exit in WCSR analysis, we adopt a similar approach with relevant constraints to achieve the WCSR. Also, we remove the PEC with its related constraints. Since WCSR also has all four optimizing variables, we apply the optimal transmit power in the iterative method to obtain the optimal UAV trajectory for WCSR. 
The attacking model presented in this paper is passive. Moreover, due to computational complexity, we have adopted the attacking model, where the adversaries' locations are static. However, it is exciting ingestion, in the case when the adversaries adjust their locations to improve the attacks. We have left this part to be completed in our future research direction. 

\item The optimization problem is non-convex.
Moreover, it is also fractional.
We apply an iterative algorithm, which considers successive convex approximation (SCA) \cite{24},
Dinkelbach \cite{25} and Taylor series expansion \cite{26}.
This iterative algorithm achieves the solution.
However, the solutions are sub-optimal.
Initially, we find optimal UAV/BS transit power control and speed under given UAV path planning and
acceleration.
Next, we minimize trajectory and acceleration using optimal power control and speed.
The algorithm repeats until convergence.
\end{itemize}

Our prior work in \cite{Sha_1} covered part of the throughput maximization considering EE UAV relaying.
We proposed an iterative algorithm by considering throughput maximization based on
los communication links and propulsion energy minimization.
On the other hand, this paper considers EE UAV relaying communication in the presence of uncertain adversaries.

The organization of our research goes as follows:
We discuss our proposed system model presented in Section II,
which defines the EE UAV relaying communications system and PEC.
We formulate EE UAV relaying maximization problem in Section III along with the sub-optimal solution.
The proposed algorithm is validated in Section IV via the simulation results.
Our work is concluded in Section V.

\section{System Model} \label{SystemModel}
\begin{figure}[!h]
\centering
\includegraphics[width=3.5in]{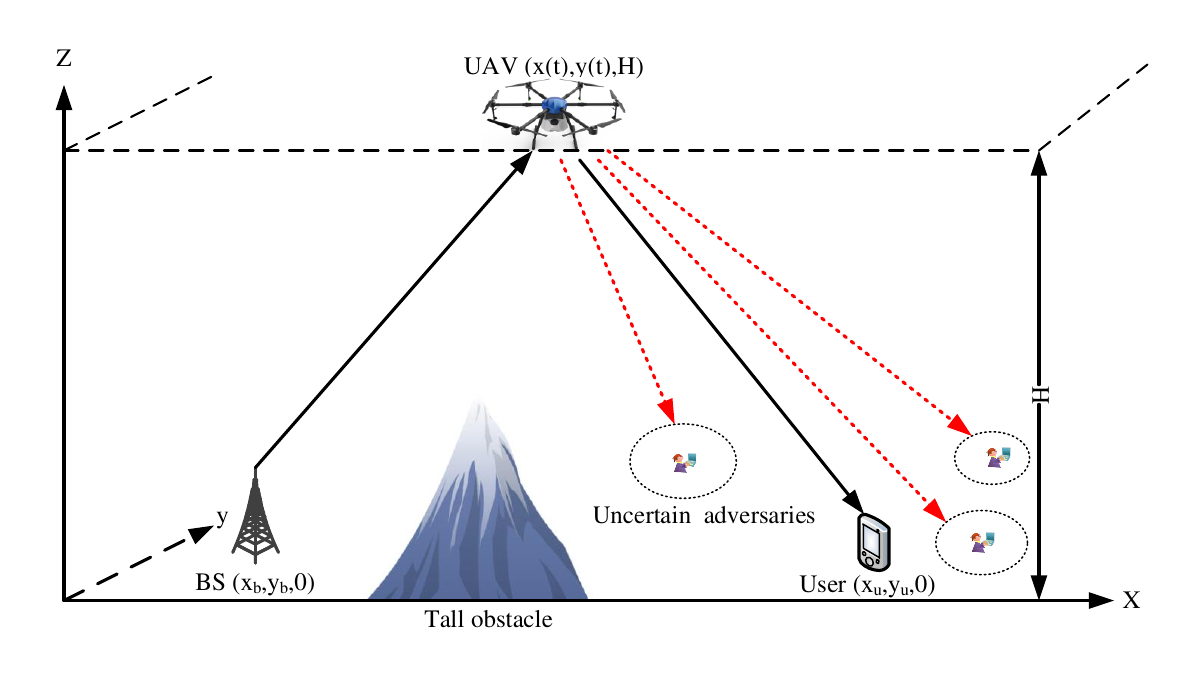}
\caption{EE UAV relay sends received information to the user at a fixed altitude, while there are the presence of uncertain adversaries and a few obstacles. On the other hand, UAV has the perfect knowledge of the BS and the user's location.}
\label{System_Model}
\end{figure}
\subsection{Description of parameters}
Fig.~\ref{System_Model} shows the information transmitted from the BS to single
a user via the UAV relay in the presence of uncertain adversaries.
We also consider a few obstacles in rural or remote areas.
Both the user and adversaries are located in this region.
No direct link is established by BS with the user and the uncertain adversaries due to long distance.
The user resides in rural or remote areas, far from BS. Due to the higher altitude of UAV and fewer obstacles, we safely assume only the los link and the system has a negligible effect on the Gaussian Additive channel, Rayleigh fading channel, or Rician fading channel. However, it is interesting to investigate the model in dense urban areas with more obstacles and los/nlos links. We have left the extension for future efforts.

Each node has a single antenna.
The UAV works as a relay that helps to communicate the BS to the user.
Thus, UAV forwards the received information from the BS to the legitimate user on the ground.
The UAV has fixed flying altitude, while both user and BS locations are known.
The UAV does not require changing its altitude since we consider the rural environment,
with a less number of higher obstacles. The UAV changes its altitude and tries to reach an optimal
altitude, which requires frequent ascending/descending.
Thus, the frequent ascending/descending will significantly consume propulsion energy,
which results that the UAV will not be able to fly a sufficiently long amount of time due to its onboard power
limitation.
The uncertain adversaries may intercept the UAV-to-user link.
In our proposed system, the UAV does not have the adversary location information.
The UAV only knows the region where the adversaries are located.
We define the uncertain adversaries set as $\cal A$=$\lbrace 1,2,3,..., A\rbrace$.
As the UAV has better los advantages, we neglect the shadowing and multi-path effects in our
proposed model.
In the few sub-sections, we explain the proposed system model in detail.

\subsection{UAV flight time and node locations} \label{IIB}
UAV provides service to the single user in $[0, T]$ time horizon, where $T$ is seconds.
Thus, $t$ is continuous. In the paper, $T$ is discretized into $N$ number of equal slots,
having slot size $\rho_t=\frac{T}{N}$ and $n=1,2,3,.....N$.
We consider that each time slot is static and equal.

We apply the BS/UAV/user/adversaries positions in a 2-D coordinate system.
Each node is static except for the time-varying UAV positions.
The time dependent UAV location is $(x[n], y[n])\in\mathbb{R}^{2\times1}$.
$H$ is defined as the UAV fixed altitude, which can avoid tall obstacles.
The users' static location is $(x_u, y_u)\in\mathbb{R}^{2\times1}$ and
the static BS location is $(x_b, y_b)\in\mathbb{R}^{2\times1}$.
The initial UAV location is $(x[2], y[2])$ due to ICC explained in Section~\ref{ICC_E}.
On the other hand the UAV final position is $(x[N], y[N],H)$.

\subsection{Uncertain adversaries}
\textcolor{black}{The adversaries can be closer to the users. In cryptographic security, it is always challenging to secure communication if the adversaries are very close to the users. Low detection probability is a viable solution to tackle the co-located adversaries and legitimate users. In the LDP, the transmission sends information to legitimate users in the presence of adversaries. However, the adversaries would not be able to detect the transmission. This can be done by changing the power level or adding a friendly jammer in the networks. Thus, it solves the case when the adversaries are close to the users. In our system model, we consider the adversaries and users, not in the exact location, and UAV can detect the two entities based on their distances. }

The UAV has both a higher los  chance of communication links and broadcast communication nature
because of higher UAV altitude.
Thus, this link sends information from the UAV to the user.
Unfortunately, due to the broadcast communication nature, the adversaries may take advantage
to intercept legitimate information.
Further, as the adversaries always obfuscate their source and destination, it is not easy for the UAV to know the actual adversarial location.

To tackle the adversary location issue, we assume UAV has the circular region of the
adversary's residence information.
The actual location of uncertain adversary $a$, where $A \in \cal A$,
is calculated from the circular region as follows:
\be
\begin{aligned}
\label{xk_1}
x_a^e=x_a+ \bigtriangleup x_a, \space \text{and}
\end{aligned}
\ee
\be
\begin{aligned}
\label{xk_3}
y_a^e=y_a+ \bigtriangleup y_a,
\end{aligned}
\ee
where $(x_a,y_a)$ is the actual adversary location $a$.
$(x_a^e, y_a^e)$ defines the estimated location of adversary $a$.
The approximated errors from actual uncertain adversary location $a$ is defined
as $(\bigtriangleup x_a, \bigtriangleup y_a) \in \varepsilon_a$,
where $\varepsilon_a$ is set of possible errors of the uncertain adversary $a$.
The following needs to be satisfied if the uncertain adversary $a$ resides on the circular region.
\be
\begin{aligned}
\label{xk2_1}
(\bigtriangleup x_a,\bigtriangleup y_a) \! \in \! \varepsilon_a \!\! \overset{\Delta}{=} \!\! \bigg \lbrace (\bigtriangleup x_a,\bigtriangleup y_a) \!\! \mid \!\! \sqrt{\bigtriangleup x_a^2 \! + \! \bigtriangleup y_a^2} \! \leq \! R_a \bigg \rbrace.
\end{aligned}
\ee
where $R_a$ is the radius of the circular region.

\textcolor{black}{The attacking model presented in this paper is passive. Moreover, due to computational complexity, we have adopted the attacking model, where the adversaries' locations are static. However, it is exciting ingestion, in the case when the adversaries adjust their locations to improve the attacks. We have left this part to be completed in our future research direction.  }

\subsection{Various links}
We use UAV flight time and node location in Section~\ref{IIB}, we can calculate the various channel gains and data rates for free space.
For example, we calculate the distance between the UAV location $(x[n],y[n],H)$ and BS location $(x_b,y_b,0)$ to achieve its corresponding channel gain. Channel gain of BS-to-UAV is:
\be
\begin{aligned}
\label{UAV_4}
g_b[n]=\frac{\beta_0}{(x[n]-x_b)^2+(y[n]-y_b)^2+H^2},
\end{aligned}
\ee
where $\beta_0$ is the channel power gain calculated when $d_0$ = 1 m (the reference distance)
\cite{QWang_1}.

At this point, we apply the Shannon capacity to achieve the rate from the formulated channel gain. The BS-to-UAV data rate is:
\be
\begin{aligned}
\label{UAV_5}
r_{b}[n] = \log_{2} \bigg( 1\! + \frac{p_b[n]g_b[n]}{\sigma^2}\bigg),\ n=1,2,...,N-1,
\end{aligned}
\ee
where $\sigma^2$ is AWGN noise power.
The BS power in $n$ time slot is $p_b[n]\in \rm I\!R_+$.
The SNR in (\ref{UAV_5}) is calculated using the channel gain between the BS and the UAV. 

Similarly, the UAV and user channel gain is expressed as follows:
\be
\begin{aligned}
\label{UAV_6}
g_u[n]=\frac{\beta_0}{(x[n]-x_u)^2+(y[n]-y_u)^2+H^2},
\end{aligned}
\ee

The UAV-to-user data rate is:
\be
\begin{aligned}
\label{UAV_7}
r_{u}[n] = \log_{2} \bigg( 1\! + \frac{p_u[n]g_u[n]}{\sigma^2}\bigg),\ n=2,3,...,N,
\end{aligned}
\ee
where the UAV power in $n$ time slot is $p_u[n]\in \rm I\!R_+$. The SNR in (\ref{UAV_7}) is calculated using the channel gain between the UAV and the user.

Similarly, channel gain of UAV-to-adversary $a$ is:
\be
\begin{aligned}
\label{UAV_50}
g_a[n]=\frac{\beta_0}{(x[n]-x_a^e)^2+(y[n]-y_a^e)^2+H^2},
\end{aligned}
\ee

The UAV-to-adversary $a$ data rate is:
\be
\begin{aligned}
\label{UAV_51}
r_a[n] = \log_{2} \bigg( 1\! + \frac{p_u[n]g_a[n]}{\sigma^2}\bigg),\ n=2,3,...,N.
\end{aligned}
\ee
The SNR in (\ref{UAV_51}) is calculated using the channel gain between the UAV and the adversaries.

\subsection{Information-causality-constraint (ICC)} \label{ICC_E}
Legitimate data is sent by BS to UAV in the time slot, $n$.
After that, UAV forwards that data to the user.
ICC states that UAV forwards the received data to the user during other time slots,
i.e. $n=2,3,4,...N$ \cite{QWang_1} is:
\be
\begin{aligned}
\label{UAV_12}
r_{u}[1]=0,
\end{aligned}
\ee
\be
\begin{aligned}
\label{C1.1}
\sum_{j=2}^{n}r_{u}[j] \leq \sum_{j=1}^{n-1}r_{b}[j], \ n=2,3,...,N.
\end{aligned}
\ee

UAV does not forward legitimate information to the user when n=1.
On the other hand, there is no transmission by the BS to the UAV when n=N.
Thus, $r_u[1]=r_e[1]=r_b[N]=0$ and $p_u[1]=p_b[N]=0$.

\begin{figure}[!h]
\centering
\includegraphics[width=3.2in]{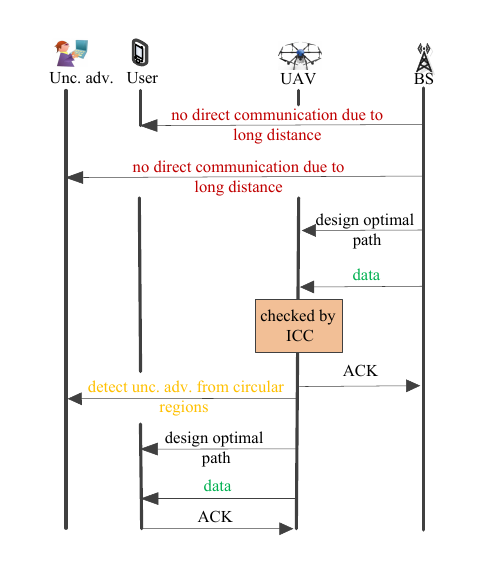}
\caption{Proposed BS-user transmission protocol via UAV relay.}
\label{TP}
\end{figure}

\subsection{UAV propulsion energy consumption (PEC)}
UAV propulsion energy is the total energy consumed for the time horizon.
It has a considerable effect on EE UAV relaying system performance.
Other energy consumption occurs due to signal processing, radiation, and electronics
circuit, etc.
This amount of energy is trivial compared to UAV propulsion energy \cite{YZeng_10}.
When the UAV has fixed wings with no abnormality, such as the backward thrust generation against the forwarding speed, the UAV hovering path becomes the propulsion energy function.
We aim to design EE UAV relaying communication via designing the optimal path planning,
velocity, acceleration, and transmit power control.
Moreover, the UAV hovering path requires an optimal trade-off to balance WCSR maximization
and energy minimization.
The UAV PEC \cite{YZeng_10} is expressed as follows:
\be
\begin{aligned}
\label{e_u_avg}
e_p[n] =\bigg (\! \alpha_u \! \parallel \!\!
v[n] \!\!\parallel^3\!+\frac{\beta_u}{\parallel\!
v[n]\!\parallel}\!+\!\frac{\beta_u \parallel \!
a[n] \!\parallel^2}{\parallel \!\!
v[n] \!\!\parallel g^2}\!\bigg) + \! \frac{\Delta k}{\rho_t}.
\end{aligned}
\ee
where both $\alpha_u$, $\beta_u$ are constant.
Their values depend on many factors, such as relay weight/wing size, etc.
$v[n]$ is the speed and $a[n]$ is acceleration.
$g$ is a gravitational constant.
Moreover, (\ref{e_u_avg}) neglects the UAV transmit power due to the meager amount of power compared
to the UAV PEC.
$\Delta k$ is kinetic energy, which is $\Delta k=\frac{1}{2}m\bigg (\parallel \! v_n[n] \! \parallel^2-\parallel \! v_n[n-1] \! \parallel^2 \bigg )$, and the mass of UAV is $m$.

\subsection{Transmission protocol}

Fig.~\ref{TP} shows a data transmission protocol for the proposed model. 
The BS sends data to the user via UAV using the optimal EE trajectory from our proposed algorithm (more detail of the proposed algorithm is explained in Section~\ref{S_3}). 
Before that, UAV verifies data by ICC and detects the uncertain adversaries from the known circular region. Finally, the UAV designs the optimal EE trajectory to communicate between the user and BS during the data transmission and UAV flight time.  Note that designing an optimal path for UAV considers the joint optimization of WCSR and PEC. The BS does not establish the user and adversaries' links throughout the process due to the user and adversaries' long distance.

\section{Optimal EE UAV relay} \label{S_3}
We design EE UAV by considering WCSR and UAV PEC.
Now we formulate EE UAV problem for UAV flight time slot, which combines (\ref{UAV_7}), (\ref{UAV_51}), and (\ref{e_u_avg}). Thus, optimization problem can be formulated with related constraints.
\begin{subequations}\label{p_1}
\begin{align}\label{ob_1.1}\
&{\mathop {\max }\limits_{x[n],y[n], p_b[n], p_u[n], v[n], a[n]}
}\ {\frac{\sum_{n=2}^N r[n]}{\sum_{n=2}^N e_p[n]}}\\
&\text{s.t.}\ \parallel v[n] \parallel \leq v_{max}, \ n=2,3,\ldots,N, \label{c_1}\\
& \parallel v[n] \parallel \geq v_{min}, \ n=2,3,\ldots,N, \label{c_2}\\
&\parallel a[n] \parallel \leq a_{max}, \ n=2,3,\ldots,N, \label{c_3}\\
& \sqrt{(x[n\!+\!1]\!-\!x[n])^2\!+\!(y[n\!+\!1]\!-\!y[n])^2}\!=\rho_t v[n] + \frac{1}{2}\rho_t^2 a[n], \label{1}\\
& 0 \leq p_b[n] \leq p_b^m, \label{UAV_8}\\
&\frac{1}{N-1}\sum_{n=1}^{N-1}p_b[n] \leq p_b^a, \label{pb_c}\\
&\frac{1}{N-1}\sum_{n=1}^{N-1}p_u[n] \leq p_u^a,\label{pu_c} \\
& 0 \leq p_u[n] \leq p_u^m, \label{UAV_10}\\
& (\ref{C1.1}). \nonumber
\end{align}
\end{subequations}
where
\be
\begin{aligned}
\label{r_1}
r[n]= r_u[n]- \max\limits_{a \in \cal A} \max\limits_{(\bigtriangleup x_a, \bigtriangleup y_a) \in \varepsilon_a}\!\!{r_a}[n].
\end{aligned}
\ee
where $r_u[n]$, $r_a[n]$, and $e_p[n]$ are expressed in (\ref{UAV_7}), (\ref{UAV_51}) and (\ref{e_u_avg}), respectively.
Moreover, (\ref{ob_1.1}) illustrates WCSR \cite{27} and UAV PEC.
(\ref{c_1}) and (\ref{c_2}) define the UAV flying speed limit.
(\ref{c_3}) is the acceleration limit.
UAV mobility expression is in (\ref{1}).
The BS peak power constraint is defined in (\ref{UAV_8}), where $p_b^m$ is the highest BS transmit power.
Average power control of the BS and the UAV are expressed in (\ref{pb_c}) and (\ref{pu_c}), respectively, where $p_u^a$ and $p_u^m$ are the UAV and BS average power.
However, (\ref{p_1}) is not easy to solve optimally due to \textit{1) not being a convex, and 2) uncertain infinite possible error numbers to find the actual adversaries locations}.
Thus, we propose an sub-optimal approach to solve (\ref{p_1}).

To solve (\ref{p_1}) sub-optimally: first, we fix the $(x[n],y[n])$ and $a[n]$ and solve $p_b[n]$, $p_u[n]$, and $v[n]$.
In the second step, we apply relay path and acceleration to achieve the optimal solution for both BS/UAV power and speed.

\subsection{Sub-optimal solution 1} \label{sb_1}
We first formulate the optimization problem to achieve $p_u[n]$, $p_b[n]$,and $v[n]$ for given $(x[n],y[n])$ and $a[n]$. Using (\ref{p_1}), the reformulated the sub-optimal problem is:
\begin{align} \label{p_2}
& \underset{p_u[n],p_b[n], v[n]}{\text{max}} \frac{\sum_{n=2}^{N} r[n]}{\sum_{n=2}^{N} e_p[n]} \\
& \text{s.t.}\ (\ref{1})~-~(\ref{UAV_10}),~(\ref{C1.1}),~(\ref{c_1}),~(\ref{c_2}).\nonumber
\end{align}

The standard optimization toolbox, such as $cvx$ cannot find the solution of (\ref{p_2}) due to the non-convexity of (\ref{C1.1}), (\ref{c_1}), (\ref{c_2}), and (\ref{p_2}).
First, we re-formulate ICC in (\ref{C1.1}) as follows:
\begin{equation}
\label{UAV_14}
\sum_{i=1}^{n-1} r_b[j] \geq \sum_{i=2}^{n} \zeta[j],
\end{equation}
\begin{equation}
\label{UAV_15}
r_u[n] \geq \zeta[n].
\end{equation}
where $\zeta[n]$ is newly added variable.
We reformulate (\ref{p_2}) as follows:
\begin{align} \label{ob_1}
& \underset{p_u[n], p_b[n], v[n],\zeta[n]}{\text{max}} \frac{\sum_{n=2}^{N} \bigg [ \zeta[n]- \max\limits_{a \in \cal A} \max\limits_{(\bigtriangleup x_a, \bigtriangleup y_a) \in \varepsilon_a}\!\!{r_a}[n] \bigg ]}{\sum_{n=2}^{N} e_p[n]}, \\
& \text{s.t.}\ (\ref{1})~-~(\ref{UAV_10}),~(\ref{c_1}),~(\ref{c_2}),~(\ref{UAV_14}),~(\ref{UAV_15}).\nonumber
\end{align}

We replace $r_u[n]$ by $\zeta[n]$, which is a new variable in (\ref{ob_1}).
However, (\ref{ob_1}) is still non-convex because of $\sum_{n=2}^{N} e_p[n]$ and (\ref{c_2}).
We apply a variable, $q[n]$ in $\sum_{n=2}^{N} e_p[n]$. Thus, the reformulated problem is expressed as follows:
\begin{subequations}\label{ob_2}
\begin{align}\label{ob_1.1.}\
&{\mathop {\max }\limits_{p_u[n], p_b[n], v[n],\zeta[n],q[n]}
}\ {\frac{\sum_{n=2}^{N} \bigg [ \zeta[n]- \max\limits_{a \in \cal A} \max\limits_{(\bigtriangleup x_a, \bigtriangleup y_a) \in \varepsilon_a}\!\!{r_a}[n] \bigg ]}{\sum_{n=2}^{N} e'_p[n]}}\\
&\text{s.t.}\ q[n] \geq v_{min}, n=2,3,\ldots,N, \label{c_4}\\
& \parallel v[n] \parallel^2 \leq q^2[n], n=2,3,\ldots,N, \label{c_5}\\
& (\ref{1})~-~(\ref{UAV_10}),~(\ref{c_1}),~(\ref{UAV_14}),~(\ref{UAV_15}).\nonumber
\end{align}
\end{subequations}
where
\be
\begin{aligned}
\label{e_u_avg_new}
e'_p[n] =\bigg (\alpha_u \parallel \!\!
v[n]\!\!\parallel^3\!+\!\frac{\beta_u}{q[n]}\!+\frac{\beta_u \parallel a[n]\parallel^2}{g^2q[n]}\!\bigg)+ \! \frac{\Delta k}{\rho_t}.
\end{aligned}
\ee

To find the solution, (\ref{ob_2}) is required to satisfy all of its constraints as a convex problem.
However, (\ref{ob_1.1.}) has still uncertain and infinite numbers of the actual locations errors of the adversaries.
To tackle the WCSR in (\ref{ob_1.1.}), following is expressed:
\begin{equation}
\label{xqwqk_2}
r_s[n]=\sum_{n=2}^N \bigg[ \zeta[n]- \log_2 (1+ \frac{p_u[n]}{\sigma^2} g_a^1[n]) \bigg ].
\end{equation}

$g_a^1$ can be defined (\ref{xk_1})~-~(\ref{xk_3}) as follows:
\begin{equation} \label{xqwqk_4}
g_a^1[n] = \frac{\beta_0}{ \min\limits_{\!\!(\bigtriangleup x_a, \bigtriangleup y_a) \in \varepsilon_a}\!\!k_a[n]}
\end{equation}

$k_a$ can be rewritten, using (\ref{xk_1})~-~(\ref{xk_3}).
\begin{equation} \label{xqwqk_5}
k_a[n] \!=\! (x_u[n]-x_a^e)^2\!+\!(y_u[n]-y_a^e)^2 \!+ \! H^2
\end{equation}

Still (\ref{xqwqk_4}) is not convex and thus, not tractable due to $(\bigtriangleup x_a, \bigtriangleup y_a) \in \varepsilon_a$. Thus, using (\ref{xk2_1}) in (\ref{xqwqk_5}), we achieve the expression.
\begin{equation} \label{xqwqk_45625634537}
\begin{split}
k_a [n]& = \! (x_u[n]\!-\!x_a)^2\!+\!(y_u[n]\!-\!y_a)^2 \!+\!H^2 \!+ \!R_a^2 \!+\!\Delta l,\\
& \approx (x_u[n]-x_a)^2+(y_u[n]-y_a)^2 +H^2 + R_a^2.
\end{split}
\end{equation}
where
\begin{equation}
\Delta l=-2\bigtriangleup x_a(x_u[n]-x_a) - 2\bigtriangleup y_a(y_u[n]-y_a).
\end{equation}

However, $(\bigtriangleup x_a,\bigtriangleup y_a)$ is very small.
However, from (\ref{xqwqk_45625634537}), the UAV-adversary $a$ distance is larger than adversary region $R_a$.
Distance between the UAV-adversary $a$ distance is:
\begin{equation}
\label{xqwsdsdq23k_4}
d_{u,a}[n]=\sqrt{(x_u[n]-x_a)^2+(y_u[n]-y_a)^2}.
\end{equation}

Following conditions can happen, such as the UAV-adversary $a$ distance is either greater/equal or less than the radius of the circular region. For example, if $d_{u,a}[n] > R_a$, then $g_a^2$ is written using (\ref{xqwqk_45625634537}) as follows:
\begin{equation}
\label{xqwq23k_4}
g_a^2[n]\! =\!\frac{\beta_0}{(\!\sqrt{(x_u[n]\!-\!x_a)^2\!+\!(y_u[n]\!-\!y_a)^2}\!-\!R_a^2)^2 \!+\!H^2},
\end{equation}

On the other hand, if $d_{u,a}[n] \leq R_a$, then $g_a^2$ is written using (\ref{xqwqk_45625634537}) as follows:
\begin{equation}
\label{2234xqwq23k_4}
g_a^2[n] =\frac{\beta_0}{H^2},
\end{equation}

We tackle the WCSR, UAV PEC, and the constraints as a convex function.
However, the EE UAV maximization problem is not yet soluble because it is still fractional in (\ref{ob_1.1.}). Due to the fractional problem, we cannot apply the optimization toolbox to achieve the solution. Thus, we employ Dinkelbach method \cite{Dinkelbach67} to tackle the objective function's fractional nature.
Fortunately, this approach confirms convergence with local optima.
\begin{align} \label{ob_3}
& \underset{p_u[n], p_b[n], v[n],\zeta[n],q[n]}{\text{max}} \sum_{n=2}^{N} \bigg [r_s[n]-\lambda_i \sum_{n=2}^{N} e'_p[n] \bigg ], \\
& \text{s.t.}\ (\ref{1})~-~(\ref{UAV_10}),~(\ref{c_1}),~(\ref{UAV_14}),~(\ref{UAV_15}),~(\ref{c_4}),~(\ref{c_5}).\nonumber
\end{align}
where $\lambda_i$ is numerical number. Moreover, $\lambda_i$ is updated in iterative fashion as $(\zeta[n]/e'_p[n])$.
Now, (\ref{ob_3}) is convex. it can be solved via convex optimization toolbox, such as $cvx$ \cite{SBoyd_1}. 

\textit{Proof.} Sub-optimal solution of (\ref{p_2}) is derived in Appendix~\ref{Proof_pu_pb}.

\textcolor{black}{We can also optimize the WCSR by removing the denominator and its related constraints as follows:
\begin{align} \label{p_28}
& \underset{p_u[n], p_b[n], \zeta[n]}{\text{max}} \sum_{n=2}^{N} r_s[n], \\
& \text{s.t.}\ (\ref{1})~-~(\ref{UAV_10}),~(\ref{c_1}),~(\ref{UAV_14}),~(\ref{UAV_15}).\nonumber
\end{align}}

\textcolor{black}{Also, (\ref{p_28}) is convex problem. We can also apply SCA to achieve the sub-optimal solution of WCSR}

\subsection{Sub-optimal solution 2} \label{sb_2}
In the subsection, we apply the solution, achieved in Section~\ref{sb_1}, to achieve the solution of the rest of the optimizing variables in our proposed model.
Using the optimal $p_u[n]$, $p_b[n]$, and $v[n]$, we reformulate the optimization problem from (\ref{p_1}) as follows:
\begin{align} \label{s_2}
& \underset{x[n],y[n], a[n]}{\text{max}} \frac{\sum_{n=2}^{N} \bigg(\Gamma_n^1-\Gamma_n^2 \bigg)}{\sum_{n=2}^{N} e_p[n]}, \\
& \text{s.t.}\ (\ref{C1.1}),~(\ref{c_3}),~(\ref{1}). \nonumber
\end{align}
where
\begin{equation}
\label{d_b^f[n]}
\Gamma_n^1=\log_2 \bigg ( 1 + \frac{p_u[n] g_u[n]}{\sigma^2}\bigg),
\end{equation}
\begin{equation}
\label{d_b^f1[n]}
\Gamma_n^2=\log_2 \bigg ( 1 + \frac{p_u[n]}{\sigma^2} \frac{\beta_0}{\min\limits_{(\bigtriangleup x_a, \bigtriangleup y_a) \in \varepsilon_a}k_a[n]} \bigg).
\end{equation}
where
\begin{equation}
k_a[n] \! = \!\! \! (x_u[n]\!-\!x_a^e)^2\!+\!(y_u[n]\!-\!y_a^e)^2 \!+\!H^2.
\end{equation}

However, (\ref{s_2}) is non-convex problem because of the fractional objective function and ICC in (\ref{C1.1}).
Due to the infinite number of $(\bigtriangleup x_a, \bigtriangleup y_a)$ possible multiple adversaries locations errors, (\ref{s_2}) is challenging to solve sub-optimally in the polynomial-time series.
We tackle non-convexity of (\ref{s_2}) by applying the slack variables $z$ and $w$.
Thus, the newly formulated optimization is:
\begin{subequations} \label{p_4.}
\begin{align}
\label{ob_4.1.}\
&{\mathop {\max }\limits_{x[n], y[n], a[n], g[n]}
}\ {\frac{\sum_{n=2}^N \bigg [ r_g[n] -r_z[n] \bigg]}{\sum_{n=2}^N e_p[n]}}\\
&\text{s.t.}\ (x[n]\! -\! x_u)^2\! +\! (y[n]\! -\! y_u)^2\! +\! H^2-g[n]\leq 0,\label{C4.1}\\
& \! \! \min\limits_{(\bigtriangleup x_a, \bigtriangleup y_a) \in \varepsilon_a} \! \! \! \! \! \! \! \!\!(x_u[n] \! - \! x_a^e)^2 \! + \! (y_u[n] \! - \! y_a^e)^2 \! + \! H^2 \! \! \geq \! \! z[n],\label{DC42}\\
& z[n] \geq H^2, \label{DC43}\\
& (\ref{C1.1}),~(\ref{c_3}),~(\ref{1}).\nonumber
\end{align}
\end{subequations}
where $\gamma=\frac{\beta_0}{\sigma^2}$, $r_g[n]=\log_{2} \bigg(1 + \frac{\gamma p_u[n]}{g[n]}\bigg)$, and $r_z[n]=\log_{2} \bigg(1 + \frac{\gamma p_u[n]}{z[n]}\bigg)$.

Thus, the similar sub-optimal solution of (\ref{s_2}) shares the similar solution of (\ref{p_4.}).
We focus on solving (\ref{p_4.}) to find the sub-optimal solution of trajectory and acceleration.

\textit{Proof.} Refer to Appendix~\ref{chap:appendixqw}.

Still (\ref{ob_4.1.}) is still a non-convex problem due to infinite possible errors from the real location $a$ in (\ref{DC42}).
Thus, we apply (\ref{xk_1})~-~(\ref{xk2_1}) in (\ref{DC42}) as follows:
\begin{equation}
\begin{aligned}
\label{rug_3}
-(x_u[n]- x_a^e)^2 -(y_u[n]- y_a^e)^2+ z[n]-H^2 \leq 0,
\end{aligned}
\end{equation}
\begin{equation}
\label{xk3_1}
\sqrt{\bigtriangleup x_a^2+ \bigtriangleup y_a^2} \leq R_a.
\end{equation}

We apply the $\cal S$-$Procedure$ mathematical approach, which can tackle the infinite number of possible uncertain location errors of adversary $a$.
Thus, a feasible point $(\bigtriangleup x_a^f, \bigtriangleup y_a^f)$ exists, for example $(\bigtriangleup x_a^f, \bigtriangleup y_a^f)=(1, 1)$, such that $\bigtriangleup {x_a^f}^2+ \bigtriangleup {y_a^f}^2 \leq R_a^2$.

The following implication also needs to be held.
\begin{equation}
\label{}
\begin{split}
-(x_u[n]- x_a - \bigtriangleup x_a)^2 -(y_u[n]- y_a - \bigtriangleup y_a)^2 \\ + z[n]-H^2 \leq 0 \Rightarrow \bigtriangleup x_a^2+ \bigtriangleup y_a^2 \leq R_a^2
\end{split}
\end{equation}

if and only if $\varepsilon_a \geq 0$ exists such that
\begin{equation}
\label{rug_2}
\left[ {\begin{array}{ccc}
\varepsilon_a[n]+1 & 0 & x_a-x_u[n] \\
0 & \varepsilon_a[n]+1 & y_a-y_u[n \\
x_a-x_u[n] & y_a-y_u[n] & m[n] \\
\end{array} } \right] \succeq 0
\end{equation}
where $m[n]=(x_u[n]-x_a)^2+(y_u[n]-y_a)^2+H^2-z[n]-R_a^2 \varepsilon_a[n]$.
Thus, (\ref{rug_2}) and (\ref{DC42}) are equivalent.
Now, (\ref{p_4.}) is:
\begin{subequations} \label{p_4}
\begin{align}
\label{ob_4.1}\
&{\mathop {\max }\limits_{x[n], y[n], a[n], g[n],z[n] \varepsilon_a[n]}
}\ {\frac{\sum_{n=2}^N \bigg [ r_g[n] -r_z[n] \bigg]}{\sum_{n=2}^N e_p[n]}}, \\
&\text{s.t.}\ \varepsilon_a[n] \geq 0, \label{DC63}\\
& z[n] \geq H^2, \label{DC43}\\
& (\ref{C1.1}),~(\ref{c_3}),~(\ref{1}),~(\ref{C4.1}),~(\ref{rug_2}).\nonumber
\end{align}
\end{subequations}
$\varphi$ is slack variables, where $\varphi \overset{\Delta}{=} [\varepsilon_1, \varepsilon_2, ......, \varepsilon_a]$, where $\varepsilon_a \overset{\Delta}{=} [\varepsilon_a[1], \varepsilon_a[2],......, \varepsilon_a[N]]^\dagger$.
Moreover, due to the non-convexity of (\ref{ob_4.1}) and (\ref{C1.1}), (\ref{p_4}) is not tractable.
On the other hand, $\log_2 ( 1 +\frac{\gamma p_u}{g[n]})$ is now convex in nature. Thus, (\ref{rug_2}) is non- linear function as it contains $[.]^2$.
Thus, non-convexity and non-linearity (\ref{p_4}) make it difficult to solve sub-optimally.
We apply an iterative algorithm, which can tackle (\ref{p_4}).
The algorithm achieves the approximate solution of (\ref{p_4}). It can be expressed as follows:The feasible sets of $(x_u,y_u, w)$ are $\mathbf{x} \overset{\Delta}{=}[x^*[1],x^*[2],......,x^*[N]]$, $\mathbf{y} \overset{\Delta}{=}[y^*[1],y^*[2],......,y^*[N]]$, and $\mathbf{w} \overset{\Delta}{=}[w^*[1],w^*[2],......,w^*[n]]$, respectively. These feasible points are also feasible to (\ref{p_4}).
Due to non-convexity of $\log_2 \bigg ( 1 +\frac{p_u^n}{w[n]} \bigg )$, we apply the first order Taylor expansion series at $w^*[n]$ as follows:
\begin{equation}
\label{xk223305_1}
\! \! \frac{p_u^n (w^*[n] \! - \! w[n])}{w^*[n](w^*[n] \! + \! p_u^n) \ln2}\!+\!\log_2 \! \! \bigg ( \! 1\!+\!\frac{p_u^n}{w^*[n]}\bigg) \! \leq \! \log_2 \! \! \bigg ( \! 1 \! + \! \frac{p_u^n}{w[n]} \! \bigg).
\end{equation}

\textcolor{black}{In (\ref{xk223305_1}), the constraint describes the relationship between the legitimate user and uncertain adversaries' rate after adopting the slack variables. We can see that the objective function in (\ref{ob_4.1}) is the non-concave function. At the same time, (\ref{rug_2}) is also not convex. This is due to the non-linearity of $(x^2[n]$,$y^2[n])$, and $m[n]$.  
We apply the first order Taylor expansion series at $z^*[n]$ for $\log_2 \bigg ( 1 +\frac{p_u^n}{z[n]} \bigg )$ as follows:}
\begin{equation}
\label{xk223405_1}
\! \! \frac{p_u^n (z^*[n] \! - \! z[n)}{z^*[n](z^*[n] \! + \! p_u^n) \ln2}+\! \log_2 \! \! \bigg ( \! 1+\frac{p_u^n}{z^*[n]}\bigg) \! \! \leq \! \log_2 \! \! \bigg ( \! 1 \! + \! \frac{p_u^n}{z[n]} \! \bigg),
\end{equation}

We tackle the non-linearity of $[.]^2$ by applying the Taylor series expansion at the feasible points.
\begin{equation}
\label{xk206_1}
-{x^*}^2[n]+2x^*[n]x[n] \leq x^2[n],
\end{equation}
\begin{equation}
\label{xk207_1}
-{y^*}^2[n]+2y^*[n]y[n] \leq y^2[n].
\end{equation}

Using (\ref{xk206_1}), (\ref{xk207_1}), we can reformulate $m[n]$ in (\ref{rug_2}) as follows:
\begin{align}
\label{xk100_1}
& m^*[n] \! = \! H^2 \! - \! z[n] \! - \! {x^*}^2[n] \! - \! {y^*}^2[n] \! + \! 2x^*[n]x[n] \! + \! x_a^2 \! + \! y_a^2 \! \! \nonumber \\ & + \! 2y^*[n]y[n] \! - \! 2x_ax[n] \! - \! 2y_ay[n] \! - \! R_a^2 \varepsilon_a[n].
\end{align}

We can transform (\ref{p_4}) using (\ref{xk223305_1})~-~(\ref{xk207_1}) as follows:
\begin{subequations} \label{OP_6}
\begin{align}
\label{DC7}\
&{\mathop {\!\!\!\! \max }\limits_{\!\!\!x_u,y_u, z, w, \varphi }
}\ \!\!\!{\frac{\!\!\sum_{n=2}^{N} \bigg [ q_1[n]-q_2[n]\bigg ]}{\sum_{n=2}^N e_p[n]} }, \\
&\text{s.t.}\ \left[ {\begin{array}{ccc}
\varepsilon_a[n]+1 & 0 & x_a-x_u[n] \\
0 & \varepsilon_a[n]+1 & y_a-y_u[n \\
x_a-x_u[n] & y_a-y_u[n] & m^*[n] \\
\end{array} } \right] \succeq 0, \label{DC68}\\
& z[n] \geq H^2, \label{DC69}\\
& \varepsilon_a[n] \geq 0, \label{DC63}\\
&(\ref{C1.1}),~(\ref{c_3}),~(\ref{1}),~(\ref{C4.1}).\nonumber
\end{align}
\end{subequations}
where
\begin{equation}
q_1[n]=\frac{p_u^n (w^*[n]-w[n])}{w^*[n](w^*[n]+p_u^n ) \ln2} -\log_2 \bigg (1 +\frac{p_u^n }{w^*[n]} \bigg )\!,
\end{equation}
\begin{equation}
q_2[n]=\frac{p_u^n (z^*[n] - z[n)}{z^*[n](z^*[n] + p_u^n) \ln2}+\log_2 \bigg ( 1+\frac{p_u^n}{z^*[n]}\bigg).
\end{equation}
where $q_1$ and $q_2$ are derived from (\ref{xk223305_1}) and (\ref{xk223405_1}), respectively.
We reformulate ICC to (\ref{UAV_14})~-~(\ref{UAV_15}) from (\ref{C1.1}).
\begin{equation}
\label{UAV_16}
\sum_{i=1}^{n-1} r_b[j] \geq \sum_{i=2}^{n} \zeta[j],
\end{equation}
\begin{equation}
\label{UAV_17}
r_u[n] \geq \zeta[n].
\end{equation}

We tackle the non-convexity of (\ref{UAV_16})~-~(\ref{UAV_17}) by adding the variable. Thus, (\ref{UAV_16}) is:
\begin{equation}
\label{UAV_18}
\sum_{i=1}^{n-1} r_b^h[j] \geq \sum_{i=2}^{n} \zeta[j],
\end{equation}
\begin{equation}
\label{UAV_19}
(x[j] - x_b)^2 + (y[j] - y_b)^2+H^2 -h[n]\leq 0.
\end{equation}
where $h[n]$ is introduced variable and
\begin{equation}
r_b^h[j]=\log_{2} \bigg( 1 + \frac{\gamma p_b[j]}{h[j]} \bigg)
\end{equation}

Now we apply Taylor series expansion at feasible point $h^{f}[j]$ in (\ref{UAV_18}).
\begin{equation}
\label{UAV_20}
\frac{\gamma p_b[j] (h^*[j] - h[j])}{h^*[j](h^*[j] +\gamma p_b[j]) \ln2}+
{r_b^h}^f[j] \leq r_b^h[j].
\end{equation}
where
\begin{equation}
{r_b^h}^f[j]=\bigg ( 1 + \frac{\gamma p_b[j]}{h^*[j]} \bigg).
\end{equation}

Thus, (\ref{UAV_16}) can be written with the help of (\ref{UAV_20}) as follows:
\begin{equation}
\label{UAV_25}
\sum_{i=2}^n \zeta[j] \leq \sum_{i=1}^{n-1}\bigg[ \frac{\gamma p_b[j] (h^*[j] - h[j])}{h^*[j](h^*[j] + \gamma p_b[j] ) \ln2}+ {r_b^h}^f[j] \bigg].
\end{equation}

Similarly, we tackle (\ref{UAV_17}) with variable $m[n]$ as follows:
\begin{equation}
\label{UAV_26}
r_u^m[n] \geq \zeta[n],
\end{equation}
\begin{equation}
\label{UAV_27}
(x[n] - x_u)^2 + (y[n] - y_u)^2+H^2 -m[n]\leq 0.
\end{equation}
where
\begin{equation}r_u^m[n]= \log_{2} \bigg( 1 + \frac{\gamma p_u[n]}{m[n]} \bigg).
\end{equation}

Now we apply Taylor series expansion at feasible point $m^{f}[n]$ in (\ref{UAV_26}).
\begin{equation}
\label{UAV_29}
\begin{split}
\frac{\gamma p_u[n] (m^*[n] - m[n])}{m^f[n](m^*[n] +\gamma p_u[n]) \ln2}+ {r_u^m}^f[n] \leq r_u^m[n],
\end{split}
\end{equation}
where
\begin{equation}
{r_u^m}^f[n]=\bigg ( 1 + \frac{\gamma p_u[n]}{m^f[n]} \bigg).
\end{equation}

Now, (\ref{UAV_17}) is reformulated using (\ref{UAV_29}) as follows:
\begin{equation}
\label{UAV_30}
\zeta[n] \leq \bigg[ \frac{\gamma p_u[n] (m^*[n] - m[n])}{m^f[n](m^*[n] + \gamma p_u[n] ) \ln2}+{r_u^m}^f[n] \bigg].
\end{equation}

Now, (\ref{p_4}) becomes:
\begin{align} \label{ob_6.1}
& \underset{x[n], y[n],\zeta[n],m[n],h[n], a[n]}{\text{max}} \! \frac{\sum_{n=2}^{N}{ \bigg [ \zeta[n]-r_z[n]\bigg ]}}{\sum_{n=2}^{N} e_p[n]}, \\
& \text{s.t.}\ (\ref{c_3}),~(\ref{1}),~(\ref{C4.1}),~(\ref{rug_2}),~(\ref{UAV_19}),~(\ref{UAV_25}),~(\ref{UAV_27}),~(\ref{UAV_30}).\nonumber
\end{align}

We impose Taylor series expansion at (\ref{ob_6.1}).
\begin{align} \label{p_51}
& \underset{x[n], y[n],\zeta[n],m[n],h[n], a[n]}{\text{max}} \! \frac{\sum_{n=2}^{N}{\bigg [\zeta[n]-O[n]\bigg]}}{\sum_{n=2}^{N} e_p[n]}, \\
&\text{s.t.}\ \varepsilon_a[n] \geq 0, \label{DC63}\\
& z[n] \geq H^2, \label{DC43}\\
& (\ref{c_3}),~(\ref{1}),~(\ref{C4.1}),~(\ref{UAV_19}), ~(\ref{UAV_25}),~(\ref{UAV_27}),~(\ref{UAV_30}).\nonumber
\end{align}
where $O[n]=\!\frac{\gamma p_u(z^*[n]\!-\!z[n])}{({z^f}^2[n]\!+\!z^f \gamma p_u[n])\ln2}$.
However, the objective function of (\ref{p_51}) is fractional problem, resulting in non-convex. We also use the classical Dinkelbach method as follows:
\begin{align} \label{p_5}
& \underset{x[n], y[n],\zeta[n],m[n],h[n], a[n]}{\text{max}} \! \! \sum_{n=2}^{N}{\! \bigg [ \! \zeta[n] \! - \! O[n]\bigg]}\! \! - \! \lambda_i \! \! \! \sum_{n=2}^{N} \! e_p[n], \\
& \text{s.t.}\ (\ref{c_3}),~(\ref{1}),~(\ref{C4.1}),~(\ref{UAV_19}), ~(\ref{UAV_25}),~(\ref{UAV_27}),~(\ref{UAV_30}).\nonumber
\end{align}

Finally, (\ref{p_5}) is a convex problem. We can apply CVX toolbox to solve (\ref{p_5}).
We combine approaches of sub-optimal solutions 1 and 2 by flitting in the iterative algorithm. Thus, we are able to solve EE UAV optimization problem in (\ref{p_1}) sub-optimally.

\textcolor{black}{We can also optimize the WCSR by removing the denominator and its related constraints as follows:
\begin{align} \label{p_9}
& \underset{x[n], y[n],\zeta[n],m[n],h[n]}{\text{max}} \! \! \sum_{n=2}^{N}{\! \bigg [ \! \zeta[n] \! - \! O[n]\bigg]}, \\
& \text{s.t.}\ (\ref{C4.1}),~(\ref{UAV_19}), ~(\ref{UAV_25}),~(\ref{UAV_27}),~(\ref{UAV_30}).\nonumber
\end{align}}

\textcolor{black}{Also, (\ref{p_9}) is convex problem. We can also apply SCA to achieve the sub-optimal solution of WCSR}

\subsection{Proposed iterative algorithm} \label{sb_3}
\textcolor{black}{We gather Sections~\ref{sb_1} and \ref{sb_2} in Algorithm~\ref{alg:algorithm_sum1} and    to solve EE UAV  relaying and WCSR optimization problem, respectively. We impose successive convex optimization for  Algorithm 1.}

Though we optimize the UAV trajectory to achieve the optimal EE, we can also investigate WCSR with the optimal UAV trajectory. Note that we have a power consumption model in the denominator of the original optimization problem. Since all the UAV trajectory variables exit in WCSR analysis, we adopt a similar approach with relevant constraints to achieve the WCSR. Also, we remove the PEC with its related constraints. Since WCSR also has all four optimizing variables, we apply the optimal transmit power in the iterative method to obtain the optimal UAV trajectory for WCSR.

\begin{algorithm}
\caption{(\ref{p_1}) solution}
\label{alg:algorithm_sum1}
\begin{algorithmic}[1]
\State $\bold{Inputting:}$ Initialize $(x_u,y_u)$, $(x_b,y_b)$, $(x_a,y_a)$ and $\gamma$.
\State Iteration, $i \longleftarrow i+1$. $(x[n],y[n])$ and $v[n]$ is updated.
\State $\bold{Initialization:}$ $x[n]={x}_{(k-1)}[n]$, $y[n]={y}_{(k-1)}[n]$, $p_u[n]={p_u}_{(k-1)}[n]$, $p_b[n]={p_b}_{(k-1)}[n]$, $v[n]=v_{(k-1)}[n]$, and $a[n]=a_{(k-1)}[n]$.
\State Compute $r_b[n]$, $r_u[n]$, $r_a[n]$, and $e_p[n]$.
\Repeat
\State Achieve optimal $p_b[n]$, $p_u[n]$, $v[n]$ using (\ref{ob_3}) for given $(x[n],y[n])$ and $a[n]$.
\State Achieve optimal $(x[n],y[n])$ and $a[n]$ using (\ref{p_5}) for the optimal solution of $p_u[n]$, $p_b[n]$, and $v[n]$.
\Until{convergence}
\end{algorithmic}
\end{algorithm}

\textcolor{black}{We summarized the optimality of the proposed algorithm as follows: Algorithm 1 solves (\ref{ob_3}) and (\ref{p_5}) alternatively and iteratively. This process continues until the solution reaches to convergence. As an example, for given $(x[n],y[n])$, and $a[n]$, the algorithm first reaches to some optimal points for $p_b[n]$, $p_u[n]$, and $v[n]$ using the feasible points. In the next iteration, using the optimal $p_b[n]$,$ p_u[n]$, and $v[n]$ from the first iteration, the algorithm searches the feasible points for $(x[n],y[n])$, and $a[n]$. In the third iteration, the algorithm again searches for $p_b[n]$, $p_u[n]$, and $v[n]$ suing the feasible points for $p_b[n]$, $p_u[n]$, and $v[n]$. This way the process goes on for large the number of iterations. Finally, the proposed algorithm guarantees the optimal solution for the optimizing variables.  
	Moreover, (\ref{UAV_29}) is lower bounded of $\log_2(1+\frac{p_u}{m^f})$. This condition can maximize the lower bound of the objective function (\ref{p_51}). The lower bound of (\ref{p_51}) is identical to the objective function of (64). However, this condition is valid only at feasible points $(x^f,y^f,m^f)$.}  

\textcolor{black}{Computational complexity of Algorithm~1 and Algorithm~2 is described as follows:
For Algorithm~1, the objective function of (\ref{p_1}) and its solutions, such as (\ref{ob_3}) and (\ref{p_5}) are increasing with the increment of iterations number. A similar trend can be obtained for Algorithm~2. 
Thus, Algorithm 1 result (\ref{p_1}) is finite.  Algorithm~\ref{alg:algorithm_sum1} is also saturated to sub-optimal solution \cite{10} having a polynomial-time solution. 
Thus, Algorithm~1 has the complexity of $\mathcal{O}[I(4N + KN)^{3.5} ]$ with $I$ number of iteration.}

\textcolor{black}{\textit{Proof}: The convergence of  Algorithm 1 is proved in Appendix C.} 

\section{Simulation Results}
We show the results, considering one user and two adversaries in the wireless networks.
We assume the user is located $(0,0)$.
The approximate locations of adversaries are $(-200,0)$ m and $(0,100)$ m.
UAV fixed altitude is 100 m.
We compare EE UAV achieved by Algorithm~\ref{alg:algorithm_sum1} to that of the scheme when the UAV works as aerial BS \cite{YZeng_10}. 
In \cite{YZeng_10}, the authors proposed an algorithm to maximize EE UAV serving a single ground user while UAV works as an aerial BS. They considered the UAV PEC.
They also considered circular trajectory, which optimized UAV flight radius and
speed to achieve the optimal EE UAV. 
In Algorithm~\ref{alg:algorithm_sum1}, we consider the location of the BS is $(650,170)$ m.
Moreover, $a=9.26 \times 10^{-4}$ and $b=2250$.
We also consider the adversaries' radii of uncertain circular regions are $R_a^1=60$ m and $R_a^2=30$ m.
The minimum UAV PEC is 100 W.
In this paper, we investigate UAV optimal trajectory and compare it to \cite{YZeng_10}.

\begin{figure*}[!t]
\normalsize
\setcounter{mytempeqncnt}{\value{equation}}
\setcounter{equation}{68}
\begin{equation}
\label{45p_u^*[34n1]}
\begin{aligned}
\begin{split}
T(p_u,p_b,v, \Gamma, \lambda_r) & = \sum_{n=2}^N \bigg [ \digamma[n] -\log_2 \bigg ( 1+\frac{p_u^*[n] g_a^2[n]}{\sigma^2} \bigg ) -\frac{g_a^2[n] (p_u^*[n]-p_u[n])}{(\sigma^2+p_u^*[n]g_a^2[n])} \bigg] \\
& +\sum_{n=2}^N \lambda_r \bigg [ \sum_{j=1}^{n-1} \log_2\bigg ( 1+ \frac{p_b[n] g_b[n]}{\sigma^2}\bigg )- \sum_{j=2}^n \Gamma [j] \bigg] \\
& -\lambda_i \bigg [ \sum_{n=2}^{N} \bigg (\alpha_u \parallel \!\!
v[n]\!\!\parallel^3\!+\!\frac{\beta_u}{q[n]}\!+\frac{\beta_u \parallel a[n]\parallel^2}{g^2q[n]}\!\bigg)+ \! \frac{\Delta k}{\rho_t} \bigg ], \\
& = \sum_{n=2}^N \bigg [ \bigg (1-\sum_{j=n}^N \lambda_j \bigg ) \digamma[n]\bigg ]- \sum_{n=2}^N \bigg [ \log_2 \bigg ( 1+\frac{p_u^*[n] g_a^2[n]}{\sigma^2} \bigg ) +\frac{g_a^2[n] (p_u^*[n]-p_u[n])}{(\sigma^2+p_u^*[n]g_a^2[n])} \bigg] \\
& +\sum_{n=1}^{N-1} \bigg [\sum_{n=1}^{N} \lambda_j \log_2\bigg ( 1+ \frac{p_b[n] g_b[n]}{\sigma^2} \bigg ]-\lambda_i \sum_{n=2}^{N} \bigg [ \bigg (\alpha_u \parallel \!\!
v[n]\!\!\parallel^3\!+\!\frac{\beta_u}{q[n]}\!+\frac{\beta_u \parallel a[n]\parallel^2}{g^2q[n]}\!\bigg)+ \! \frac{\Delta k}{\rho_t} \bigg ].
\end{split}
\end{aligned}
\end{equation}
\hrulefill
\vspace*{4pt}
\end{figure*}

We explain the comparison of different parameters, such as UAV speed, acceleration, user and adversaries rate, propulsion power consumption, and EE UAV of different algorithms in Table~\ref{tab:my-table}.
\begin{table*}[]
\caption{Various EE UAV Algorithms}
\label{tab:my-table}
\centering
\begin{tabular}{|c|c|c|c|c|c|}
\hline
&\begin{tabular}[c]{@{}c@{}}Average \\ speed ($m/s$)\end{tabular} & \begin{tabular}[c]{@{}c@{}}Average \\ acceleration ($m/s^2$)\end{tabular}& \begin{tabular}[c]{@{}c@{}}Average user\\ \& Adversary rate $(Mbps$)\end{tabular}& \begin{tabular}[c]{@{}c@{}}Average \\ power ($W$)\end{tabular} & \begin{tabular}[c]{@{}c@{}}EE UAV\\ ($Kbits/J$)\end{tabular} \\ \hline
Algorithm 1 & 26.23 & 3.01 & 10.12 \& 1.22 & 116.98 & 74.55 \\ \hline
EE UAV \cite{YZeng_10} & 25.67 & 3.24 & 8.34 \& 0 & 116.02 & 71.89 \\ \hline
EE UAV, circular \cite{YZeng_10}& 25.20 & 4.02 & 8.16 \& 0 & 119.10 & 68.56 \\ \hline
\end{tabular}
\end{table*}

\begin{figure}[!h]
\centering
\includegraphics[width=3.6in]{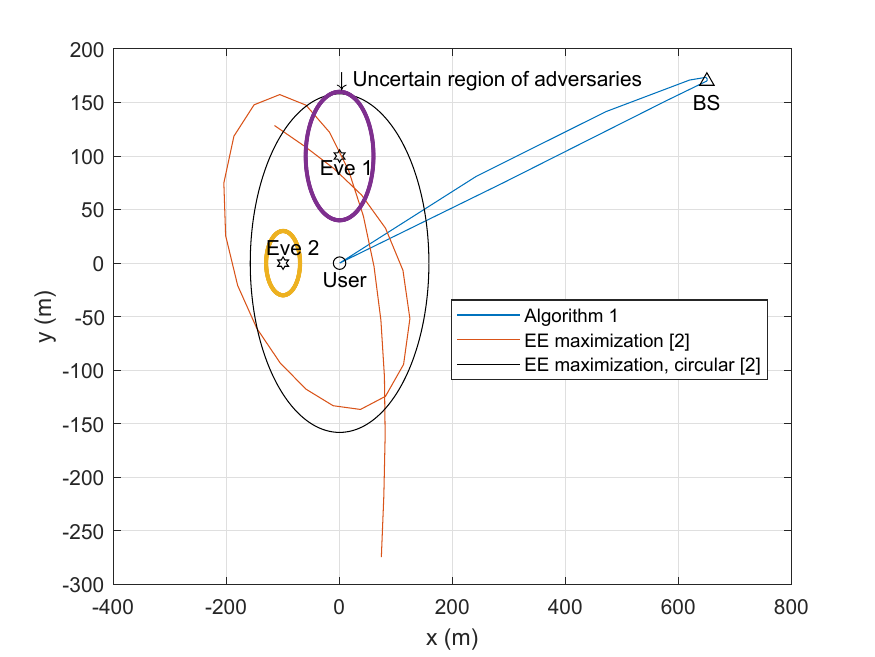}
\caption{The UAV optimal trajectory design.}
\label{XY}
\end{figure}

Fig.~\ref{XY} shows the UAV optimal hovering path for the proposed Algorithm~\ref{alg:algorithm_sum1} and compares with \cite{YZeng_10}.
For Algorithm~\ref{alg:algorithm_sum1}, the UAV flies between the BS and user.
During the entire flight, the UAV keeps a safe distance from the adversaries though the UAV does have the perfect location knowledge of the adversaries.
The optimal hovering path is narrower and more directive compared to \cite{YZeng_10}.
Though the UAV serves as an aerial BS in \cite{YZeng_10}, their proposed algorithm shows a wider optimal path.
As a result, the UAV consumes more energy compared to Algorithm~\ref{alg:algorithm_sum1}.
Moreover, Algorithm~\ref{alg:algorithm_sum1} shows a better optimal path, which can save UAV PEC while the UAV is working as a relay.
\begin{figure}[!h]
\centering
\includegraphics[width=3.6in]{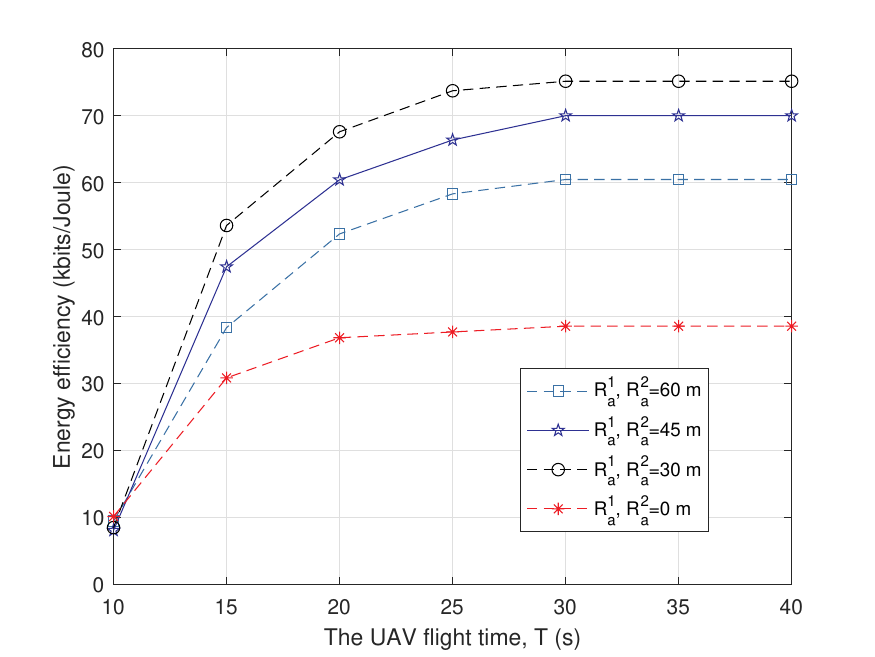}
\caption{Effect of the uncertain circular region of the adversaries on energy efficiency.}
\label{LM}
\end{figure}

Fig.~\ref{LM} shows EE UAV for the proposed Algorithm~\ref{alg:algorithm_sum1} when the radii of adversaries uncertain circular are 60 m, 45 m, and 30 m, respectively. This is then compared with the non-robust scheme, which defines the UAV considers the approximated locations as the exact location, i.e., $R_a = 0$. 
We consider the same radius for both adversaries in our investigations.
As shown in Fig.~\ref{LM}, the high EE UAV Algorithm 1 is achieved when the circular region is more significant due to the rate maximization.
On the other hand, EE UAV significantly drops when the UAV considers the approximated locations as the perfect locations.
\begin{figure}[!h]
\centering
\includegraphics[width=3.6in]{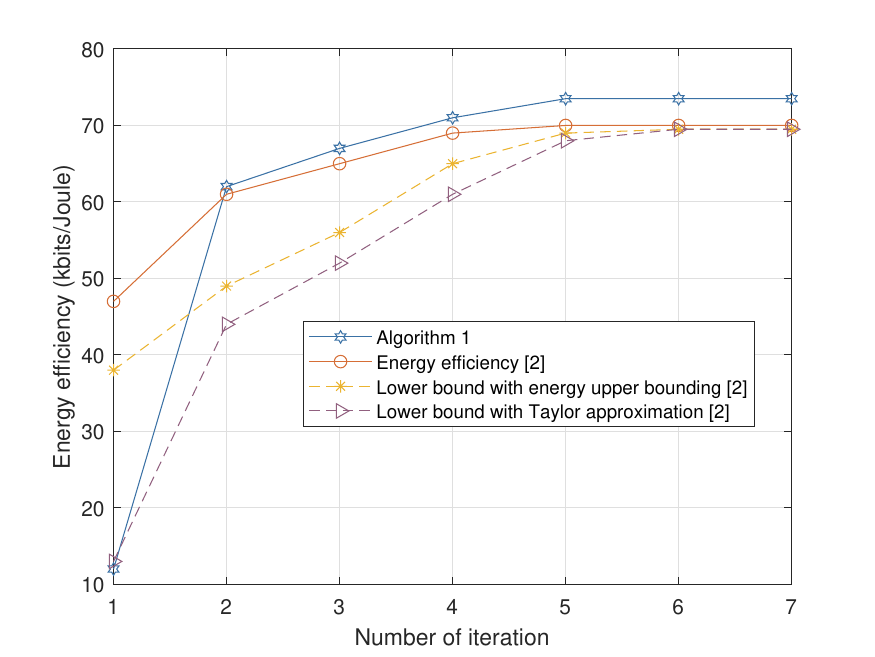}
\caption{Convergence of algorithms.}
\label{EE}
\end{figure}

Algorithm 1 convergence is demonstrated in Fig.~\ref{EE}. Fig.~\ref{EE} consists of several curves, such as Algorithm~1, EE UAV defining EE UAV based on PEC model in \cite{YZeng_10}, lower bound with energy upper bounding defining the energy upper bounding given in \cite{YZeng_10}, and lower bound with Taylor approximation defining the local convex approximation \cite{YZeng_10}. Fig.~\ref{EE} shows that Algorithm 1 monotonically converges. Furthermore, it shows that the adopted lower bounds for efficient convex optimization are rather tight, primarily due to the convergence algorithm.

\begin{figure}[!h]
\centering
\includegraphics[width=3.6in]{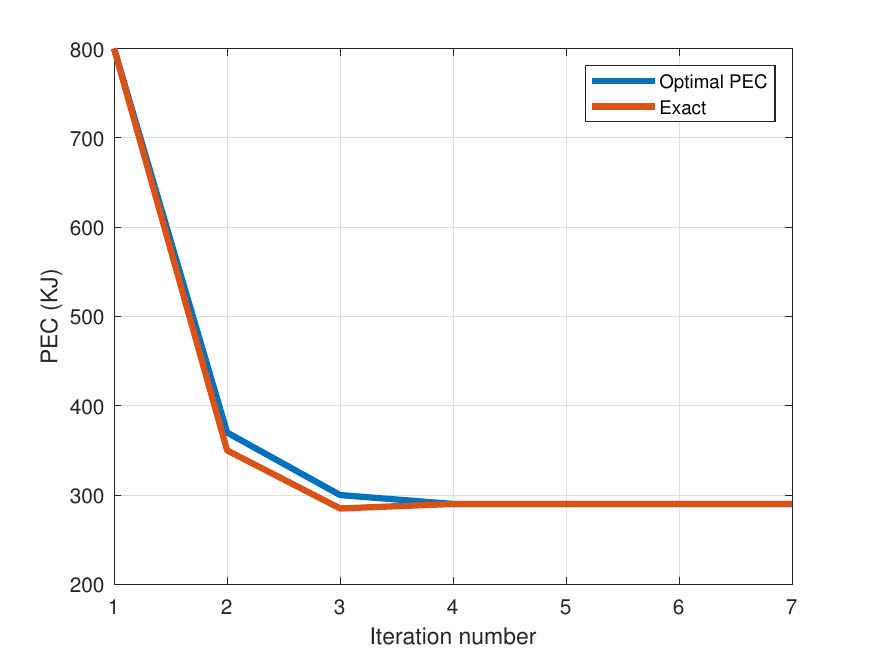}
\caption{Convergence of PEC minimization.}
\label{PEC}
\end{figure}

Fig.~\ref{PEC} shows the convergence of PEC from our proposed Algorithm~1. We define ``Exact" as the energy consumption calculated based on (\ref{e_u_avg}).
The optimal PEC is very close to the value of "Exact." We claim that the proposed PEC minimization is practically tight, given that we solve it via convex optimization.
Fig.~\ref{e_u_avg} also shows that PEC minimization converges after iteration. 
Thus, the efficiency of Algorithm~1 is proved to be efficient.

\begin{figure*}
\centering
\label{P_Fig}
\begin{subfigure}[b]{0.49\textwidth}
\centering
\includegraphics[width=\textwidth]{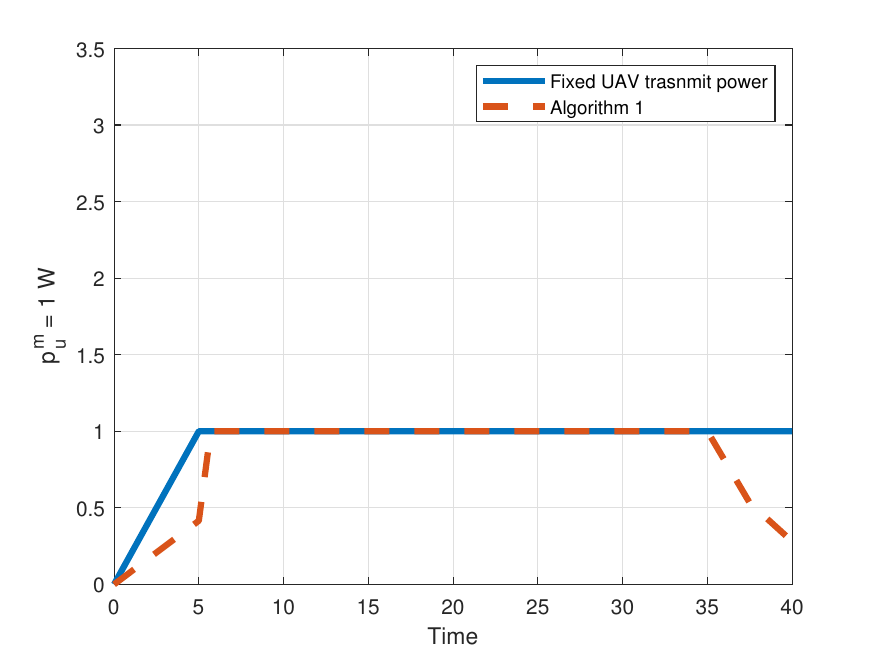}
\caption{$p_u^{m}=1$ W}
\label{P_1}
\end{subfigure}
\hfill
\begin{subfigure}[b]{0.49\textwidth}
\centering
\includegraphics[width=\textwidth]{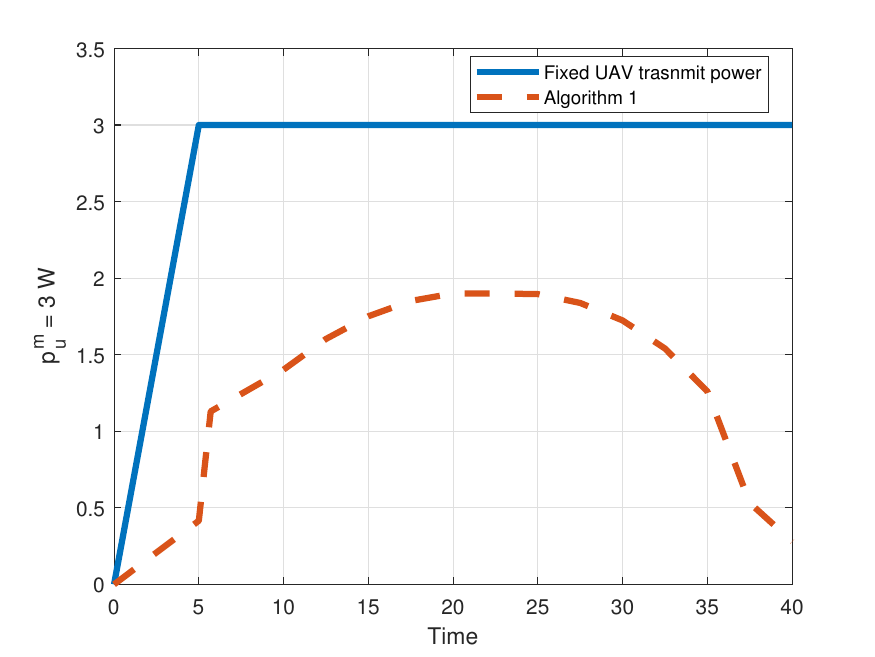}
\caption{$p_u^{m}=3$ W}
\label{P_3}
\end{subfigure}
\caption{UAV transmit power for flight time.}
\label{power}
\end{figure*}
Fig.~\ref{power} shows the UAV transmits power to change for UAV flight time. 
There exists a trade-off between transmit power and EE UAV. The UAV transmits higher power that may disclose secure information to the adversaries. We keep the BS transmit power fixed and vary UAV transmit power. When the transmit power is low, Algorithm~1 results in a saturated power level, as shown in Fig.~\ref{P_1}. However, Fig.~\ref{P_3} shows the required $p_u$ and shows the power savings for Algorithm 1.

\begin{figure}[!h]
\centering
\includegraphics[width=3.6in]{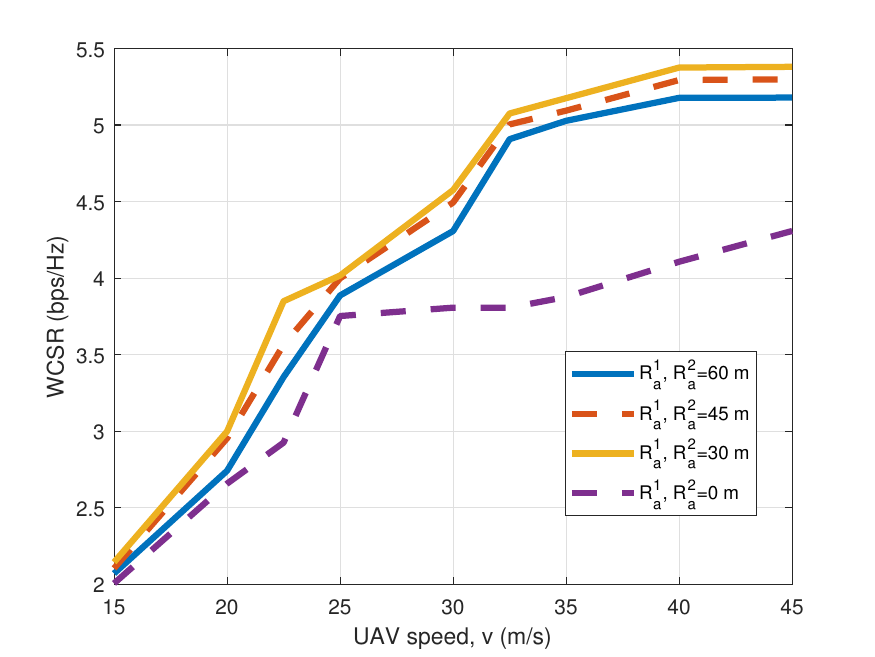}
\caption{Effect of the uncertain circular region of the adversaries on WCSR for different velocities.}
\label{th}
\end{figure}

Fig.~\ref{th} shows WCSR when the radii of adversaries uncertain circular are 60 m, 45 m, and 30 m, respectively. Then, this is compared with the non-robust scheme, i.e., $R_a = 0$. 
\textcolor{black}{WCSR saturated to sub-optimal solution \cite{10} having a polynomial-time solution. 
WCSR is achived by solving  (\ref{p_28}) and (\ref{p_9})  iteratively and alternatively. 
A similar change is obtained with the complexity of $\mathcal{O}[I(4N + N)^{3.5} ]$ with $I$ number of iteration. WCSR reaches convergence.}
We consider the same radius for both adversaries in our investigations.
As shown in Fig.~\ref{th}, the high WCSR Algorithm 1 is achieved when the circular region is more significant due to the rate maximization.
On the other hand, WCSR significantly drops when the UAV considers the approximated locations as the perfect locations.

\section{Conclusions}
We consider the UAV-assisted wireless networks that jointly optimize the UAV trajectory, speed, acceleration, and UAV/BS power in the presence of uncertain adversaries.
The UAV works as a mobile relay, which transmits information from BS to the user.
ICC is imposed to make sure that the UAV sends the received information to the user.
We formulate the WCSR maximization problem while the environment has uncertain adversaries.
We also apply the UAV PEC, which eventually leads to designing EE UAV communications.
We propose an algorithm, which tackles the optimization problem sub-optimally.
Though our algorithm is sub-optimal, it is computationally solvable.
First, the UAV/BS power and UAV speed are optimized for the given UAV trajectory, UAV acceleration.
Then, the use of the UAV/BS transmit power and UAV speed, the UAV trajectory, and UAV acceleration are optimized.
Then, this system is repeated until convergence.
We present the simulation outcomes based on the proposed Algorithm~\ref{alg:algorithm_sum1} that shows the UAV hovers between the users and the BS to assure ICC is applied.

\appendices
\section{Power Control and Speed Solution}\label{Proof_pu_pb}
The $p_b[n]$ and $p_u[n]$ optimal solution can be explained as form of liquid filling.
$p_b[n]$ and $p_u[n]$ have different liquid level.
The BS power control is staircase liquid filling, considering $\flat$ is both non-negative and non-increasing.
Due to uncertain adversaries, the UAV transmit power liquid level does not experience the same monotone over the time horizon.
We aim to achieve the dual optimal variables, $\lbrace \lambda_r \rbrace_{n=2}^N$. Note that $\lbrace \lambda_r \rbrace_{n=2}^N$ optimizes Lagrange dual function.
We apply the ellipsoid method \cite{40} for dealing with the Lagrange duality.
\be
\begin{aligned}
\label{159wd}
\sum_{n=2}^N \lambda_r \leq 1,
\end{aligned}
\ee
\be
\begin{aligned}
\label{45p_u^*[n0]}
\lambda_r \geq 0.
\end{aligned}
\ee

We see that \ref{159wd}~-~\ref{45p_u^*[n0]} can minimize Lagrange dual function.
By applying the above methods, the EE UAV problem can be maximized.
$p_u[n]$ updates in an iterative fashion while achieving the EE UAV maximization problem.
The optimal solution of UAV and BS power control can be proved in another way.


The partial Lagrangian function is $T(p_u,p_b,v, \Gamma, \lambda_r)$.
The $T(p_u,p_b,v, \Gamma, \lambda_r)$ is expressed in (\ref{45p_u^*[34n1]}).
Thus, the Lagrange dual function for (\ref{ob_3}) is:
\begin{subequations} \label{OP_9}
\begin{align}
\label{DC3}\
& {\mathop {\max }\limits_{p_u[n], p_b[n], v[n], \Gamma, \lambda_r}
}\ T(p_u,p_b, v, \Gamma, \lambda_r), \\
&\text{s.t.}\ (\ref{1})~-~(\ref{UAV_10}),~(\ref{c_1}),~(\ref{UAV_14}),~(\ref{UAV_15}),~(\ref{c_4}),~(\ref{c_5}) \nonumber.
\end{align}
\end{subequations}

We focus on (\ref{OP_9}), which is maximized using (\ref{45p_u^*[34n1]}). It also achieves the dual function, having $\lambda_r$, where the solution is achieved by minimizing the dual function.
$\lambda_r^*$ determines UAV transmit power, and BS transmit optimal power solution.
Thus, the minimization of Lagrange dual function minimization over UAV and BS power, considering $\lambda_r$ is fixed, can be achieved from (\ref{45p_u^*[34n1]}).

${f_p}_{b}(\lambda_r)$ defines the BS power function and ${f_p}_{u}(\lambda_r)$ defines the UAV power function. The function over the UAV speed if ${f_v}(\lambda_r)$. The following expression can be written as follows:
\begin{equation}
\label{415p_b^*[n1]}
f(\lambda_r)={f_p}_{u}(\lambda_r)+{f_p}_{b}(\lambda_r)+{f_v}(\lambda_r).
\end{equation}
where
\begin{subequations} \label{OP_10}
\begin{align}
\label{DC10}\
& {f_p}_{u}(\lambda_r)\!=\!{\mathop {\!\! \max }\limits_{p_u[n], \Gamma [n]}
}\ \!\! \sum_{n=2}^N \! \digamma[n]\!-\! \log_2 \! \bigg ( \!1 \!+ \!\frac{p_u^*[n] g_a^2[n]}{\sigma^2} \bigg )\!- \!\eta_n , \\
&\text{s.t.}\ (\ref{UAV_10}),~(\ref{UAV_14}),~(\ref{UAV_15})\nonumber.
\end{align}
\end{subequations}
where
\begin{equation}
\label{415p_b^*[n1]}
\eta_n=\frac{g_a^2[n] (p_u^*[n]-p_u[n])}{(\sigma^2+p_u^*[n]g_a^2[n])}.
\end{equation}

${f_p}_{b}(\lambda_r)$ is:
\begin{subequations} \label{OP_11}
\begin{align}
\label{D2515151C11}\
& {f_p}_{b}(\lambda_r)={\mathop {\max }\limits_{p_b[n]}
}\ \!\!\!\sum_{n=1}^{N-1} (\sum_{n=1}^{N} \lambda_j)\log_2\bigg ( 1+ \frac{p_b[n] g_b[n]}{\sigma^2} \bigg ), \\
&\text{s.t.}\ (\ref{UAV_8}),~(\ref{UAV_14}),~(\ref{UAV_15})\nonumber.
\end{align}
\end{subequations}

Moreover, ${f_v}(\lambda_r)$ is:
\begin{subequations} \label{OP_11}
\begin{align}
\label{D2v}\
& {f_v}(\lambda_r)={\mathop {\max }\limits_{v[n]}
}\ \!\!-\!\!\lambda_i \!\! \sum_{n=2}^{N} \!\! \bigg [\! \bigg ( \!\! \alpha_u \parallel \!\!
v[n]\!\!\parallel^3\!+\!\frac{\beta_u g^2\!\! \parallel \!\! a[n] \!\! \parallel^2}{g^2q[n]}\!\bigg) \!\!+ \! \frac{\Delta k}{\rho_t} \bigg] \\
&\text{s.t.}\ (\ref{1}),~(\ref{c_1}),~(\ref{c_4}),~(\ref{c_5}) \nonumber.
\end{align}
\end{subequations}

Using (\ref{OP_10})~-~(\ref{OP_11}), we can obtain the sub-optimal the UAV/BS transmit power and speed.
Since, Lagrange dual variable, $\lambda_r$ is given while solving those (\ref{OP_10})~-~ (\ref{OP_11}), we employ the standard Karush Kuhn Tucker conditions. Eventually, the optimal UAV/BS power and the speed are achieved.
The proof is now complete.

\section{Shared UAV Trajectory and Acceleration Solution}\label{chap:appendixqw}
The following expressions are active.
\begin{equation}
\label{45p_b^*[n12121]}
x^2[n]+y^2+H^2-w[n] \leq 0,
\end{equation}
\begin{equation}
\label{45p_u^*[n231]}
\min\limits_{\!\!(\bigtriangleup x_a, \bigtriangleup y_a) \in \varepsilon_a}\!\!(x[n]\!-\!x_a^e)^2\!+\!(y[n]\!-\!y_a^e)^2 \!+\!H^2 \!\geq z[n].
\end{equation}

We can prove that the sub-optimal solution of (\ref{p_4}) and (\ref{s_2}) is equivalent by the theory of contradiction.
If (\ref{45p_b^*[n12121]}) and (\ref{45p_u^*[n231]}) are not active, the objective function of (\ref{p_4}) is $\frac{\sum_{n=2}^N \bigg [ r_g[n] -r_z[n] \bigg]}{\sum_{n=2}^N e_p[n]}$.
It improves with the increment of $z[n]$.
The decrements of $w[n]$ improve the objective function.
The proof is now complete.

\section{Proof of the convergence of the algorithm}\label{chap:appendixqw}
\textcolor{black}{We can prove the convergence by recursive formula. Lets define the following expression:}
\begin{align}
\label{AppC1}
\textcolor{black}{\frac{\tau-\tau^{(t+1)}}{\tau-\tau^{(t)}}=T, m \geq 0,}
\end{align}

\textcolor{black}{Lets replace the $t$ with $(t-1)$. We can rewrite (\ref{AppC1}) as follows:}
\begin{align}
\label{AppC2}
\textcolor{black}{\frac{\tau-\tau^{(t)}}{\tau-\tau^{(t-1)}}=T, m \geq 1,}
\end{align}

\textcolor{black}{By subtracting (\ref{AppC2}) from (\ref{AppC1}), we get}
\begin{align}
\label{AppC3}
\textcolor{black}{\frac{\tau^{(t+1)}-\tau^{(t)}}{\tau^{(t)}-\tau^{(t-1)}}=T, m \geq 1,}
\end{align}

\textcolor{black}{In the case, when $|| T ||$, we can rewrite (\ref{AppC3}) as follows:}
\begin{align}
\label{AppC4}
\textcolor{black}{\frac{|| \tau^{(t+1)}-\tau^{(t)}||}{||\tau^{(t)}-\tau^{(t-1)}||} \geq ||T||},
\end{align}

\textcolor{black}{We estimate $c = M$ by using
the ratio, in other words, the successive differences of the norms.
In most cases, these ratios are approximately
constant with the increment of $\tau$. Thus, we can write:}
\begin{align}
\label{AppC4}
\textcolor{black}{{c_t= \mathop {\max }\limits_{{t-p \geq q \geq t}}
}  \frac{||\tau^{(q+1)}-\tau^{(q)}||}{\tau^{(q)}-\tau^{(q-1)}}}
\end{align}

\textcolor{black}{Finally, we can write:}
\begin{align}
\label{AppC5}
\textcolor{black}{\frac{|| \tau-\tau^{(m+1)}||}{||\tau^{(m+1)}-\tau^{(m)}||} \geq \frac{c}{1-c}}
\end{align}

\textcolor{black}{Thus, the proof is completed.}

\ifCLASSOPTIONcaptionsoff
\newpage
\fi

\end{document}